\documentclass[aps,prb,twocolumn,showpacs,amsmath,amssymb]{revtex4}
\usepackage{CJK}
\usepackage{txfonts}
\usepackage{color}
\usepackage{graphicx}
\usepackage{cases}
\usepackage{dcolumn}
\usepackage{bm}

\begin{document}
\begin{CJK*}{GB}{gbsn}
\title{Electronic and transport properties of azobenzene monolayer junctions as molecular switches}

\author{Yan Wang (Íõçü)}
\author{Hai-Ping Cheng}
\affiliation{Department of Physics and Quantum Theory Project, University of Florida, Gainesville, Florida
32611, USA}

\begin{abstract}
We investigate from first-principles the change in transport properties of a two-dimensional azobenzene
monolayer sandwiched between two Au electrodes that undergoes molecular switching. We focus on transport
differences between a chemisorbed and physisorbed top monolayer-electrode contact. The conductance of the
monolayer junction with a chemisorbed top contact is higher in \textit{trans} configuration, in agreement
with the previous theoretical predictions of one-dimensional single molecule junctions. However, with a
physisorbed top contact, the "ON" state with larger conductance is associated with the \textit{cis}
configuration due to a reduced effective tunneling pathway by switching from \textit{trans} to \textit{cis},
which successfully explains recently experimental measurements of azobenzene monolayer junctions. A simple
model is developed to explain electron transmission across subsystems in the molecular junction. We also
discuss the effects of monolayer packing density, molecule tilt angle, and contact geometry on the calculated
transmission functions. In particular, we find that a tip-like contact with chemisorption significantly
affects the electric current through the \textit{cis} monolayer, leading to highly asymmetric current-voltage
characteristics as well as large negative differential resistance behavior.

\end{abstract}


\maketitle
\end{CJK*}

\section{\label{sec:1}INTRODUCTION}
Molecular junctions that incorporate photochromic molecules as reversible photo-switches between two
different conductance states ("ON" and "OFF") have advanced considerably in the last
decade\cite{Russew:2009,Molen:2010}. Azobenzene and its derivatives are the most frequently studied
candidates for photoresponsive molecular switch, based on their conformational changes from a more
thermodynamically stable \textit{trans} configuration to a \textit{cis} configuration in response to an
external stimulus such as UV light\cite{Chang:2004}, and vice versa upon exposure to visible light or thermal
excitation\cite{Ikegami:2003}. Previous first-principles studies for azobenzene molecular junctions focus on
electron transport through ideal single-molecule junctions with one-dimensional (1D) electrodes and predict
that the junction with \textit{trans} configuration has a conductance higher than that with the \textit{cis}
configuration \cite{Zhang:2004,Zhang:2006,Valle:2007}.

More recent attention in experiments has been paid to the self-assembled monolayer (SAM) systems, which are
highly ordered arrays of molecules on a two-dimensional (2D) surface with a chemisorbed bottom contact. A
variety of experimental methods are used to apply a second contact on top of the azobenzene SAM and measure
the photoinduced changes in its conductance, with sufficient flexibility to adapt to the height change of the
SAM after its photo-switch, including liquid metal contact (eg., Hg drop \cite{Ferri:2008}) and conducting
atomic force microscopy \cite{Mativetsky:2008,Smaali:2010}. These pioneering experiments found that the "ON"
state with larger measured current is associated with the \textit{cis} configuration, contrasting previous
theoretical predictions based on single-molecule junction models.

These experiments using 2D monolayer junctions present more complicated situations in which the ideal 1D
single-molecule junction model may not apply. First, in the monolayer junction the two monolayer-electrode
contacts are not symmetric. The one at the bottom interface is a chemisorbed contact, in which the end group
of the molecule is chemically bonded to the electrode. The top contact is merely physical, i.e., no chemical
bonding formed between the SAM and the top electrode. It is known that the way the molecule interacts with
the electrode plays a crucial role in electronic transport through the molecule. The conductance of a
junction with and without chemicontacts can differ by several orders of magnitude
\cite{Cui:2001,Taylor:2002}. In addition, SAMs can form with differences in packing density and tilt angle,
and in densely packed SAMs the conductance may involve both intramolecular and intermolecular contribution
due to molecules in close proximity \cite{Agapito:2008}. Other effects such as the surface topography and
roughness of the contact also influence the final geometry of the contact and thus lead to more complex
transport properties. These uncertainties in the microscopic details of a monolayer junction make
understanding the effects of monolayer-electrode coupling, intermolecular interaction and contact geometries
critical to elucidating its transport mechanisms. First-principles studies are especially valuable in this
regard because information concerning the microscopic details of a monolayer junction and their effects on
electron transport are often impossible to obtain directly in experiments.

Here we report a first-principles study of the electronic structure and transport properties of azobenzene
monolayer junctions. The monolayer junction is constructed by attaching a Au(111) electrode on top of an
azobenzene monolayer chemisorbed on Au(111) substrate. We focus on clarifying the role of monolayer-electrode
bonding in relation to the transport properties by investigating the junctions with two types of top
monolayer-electrode contact: 1) a strongly bonded contact in which the molecules are covalently bonded to the
Au surface, and 2) a weakly bonded contact which is physisorbed through van der Waals (vdW) interation. We
find that both the \textit{trans} to \textit{cis} transformation of the molecules and the monolayer-electrode
bonding play critical roles in determining the electron transmission as well as the current-voltage
characteristics of the monolayer junction. The impact of monolayer packing density and tilt angle are
discussed. We also consider the effect of contact geometry by adding an additional Au atom at the top contact
and find substantive changes in transport depending on this particular contact geometry.

\section{\label{sec:2}computational methods}

We use two density functional theory (DFT) based computational methods in our calculations. First, all
structural optimizations, including calculations for both the azobenzene monolayer on a Au(111) surface and
the molecular junction consisting of two monolayer-electrode contacts, are performed using the
plane-wave-basis-set Vienna \textit{ab} initio simulation package (VASP) \cite{VASP:1996}. Then, electron
transport calculations of the molecular junction are performed using the atomic-orbital-basis-set TRANSIESTA
code \cite{Brandbyge:2002}, which incorporates nonequilibrium Green's function formalism \cite{NEGF} and DFT
treatment of the electronic structure as implemented in the SIESTA package \cite{Soler:2002}.

In the structural optimization calculations, projector augmented wave potentials with kinetic energy cutoff
of 400 eV are employed. For the exchange and correlation functional we utilize the Perdew-Burke-Ernzerh
generalized gradient approximation (PBE-GGA) \cite{Perdew:1996}. A $3\times3\times1$ \textit{k}-point
sampling is applied based on the Monkhorst-Pack \cite{Monkhorst:1976} scheme. All geometries are optimized
until the remaining force on each ion fall below the convergence criterion of 0.02 eV/{\AA}.

In the electron transport calculations, we utilize norm-conserving nonlocal Troullier-Martins
pseudopotentials and the PBE-GGA exchange correlation functional, with a single-$\zeta$ plus polarization
basis for the Au atoms and a higher level double-$\zeta$ basis set for the molecules. The equivalent
plane-wave cutoff for the real-space grid of 150 Ry is used throughout the calculations. A $5\times5\times1$
Monkhorst-Pack \textit{k}-point sampling is used for self-consistent calculation, and a $9\times9$
\textit{k}-point sampling is used for transmission calculations. The transmission function is obtained by
averaging transmission coefficients in each \textit{k}-point for an applied bias voltage $V$,
$T(E,V)=\sum_{k_x,k_y}T_{k_x,k_y}(E,V)$, where
\begin{equation}
T_{k_x,k_y}={\rm Tr}[\Gamma_1 G^R\Gamma_2 G^A]
\end{equation}
is the \textit{k}-resolved transmission coefficient, $G^{R,A}$ are the retarded (advanced) Green's function
matrices, and $\Gamma_{1,2}=i[\Sigma_{1,2}-\Sigma_{1,2}^\dag]$ where the $\Sigma_{1,2}$ are the retarded
self-energies due to the existence of bottom (top) semi-infinite electrodes. The electric current as a
function of the applied voltage is obtained by the integration of the transmission function
\begin{equation}
I(V)=\frac{2e}{h}\int T(E,V)[f(E-\mu_1)-f(E-\mu_2)] dE.
\end{equation}
where $e$ is the elementary charge, $f$ is the Fermi-Dirac distribution function, $\mu_{1,2}=E_{\rm F} \pm
eV/2$ are the chemical potentials for the bottom (top) electrodes.

\section{\label{sec:3}Azobenzene molecular monolayer on Au(111) surface}

\subsection{\label{sec:31}Structural models}

\begin{figure}
{\includegraphics[width=7cm]{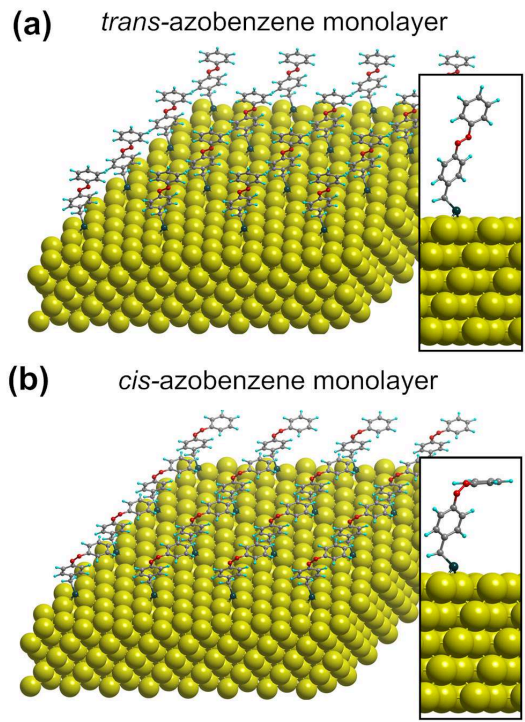}} \caption{\label{fig:monolayer}(Color online). Equilibrium
configurations of \textit{trans} (a) and \textit{cis} (b) azobenzene molecules chemisorbed on Au(111) surface
via a $\rm CH_2S$ linker. Insets: side view of the corresponding geometry. C, H, N, S and Au atoms are
colored gray, green, red, blue and yellow, respectively.}
\end{figure}

For the azobenzene monolayer on the Au(111) substrate (hereafter we denote this as Au-AB), the systems that
we study consist of an azobenzene monolayer chemisorbed on the surface of Au(111) substrate by a $\rm CH_2S$
linker for each molecule, as illustrated in Fig.\ \ref{fig:monolayer} produced by VESTA graphical program
\cite{Vesta3}. The substrate is modeled by a slab consisting of five Au(111) layers. The molecular coverage
is kept fixed at one molecule per $(\sqrt{3}\times\sqrt{3})R30\textordmasculine$ surface unit cell, which
corresponds to a packing area of 90.8 $\AA$ per molecule. The supercell thus consists of 87 atoms comprising
12 atoms per Au layer and 27 atoms for the azobenzene molecule with the linker. A large vacuum spacing of 15
{\AA} is used between the topmost molecular atom and the next Au slab to prevent interaction between adjacent
images. All atoms except the two bottom Au layers are fully relaxed.

\subsection{\label{sec:32}Adsorption geometry and energetics}

\begin{figure}
{\includegraphics[width=8cm]{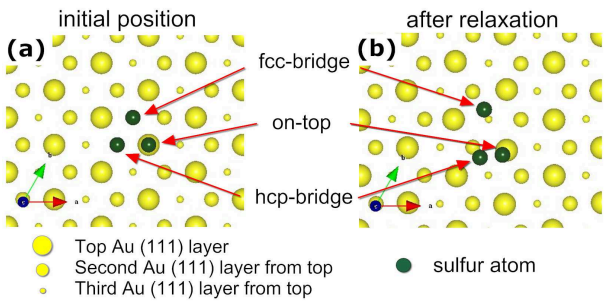}} \caption{\label{fig:sites}(Color online). (a) Top view
indicating the lateral position of the sulfur atom in the structural optimizations carried out to determine
the most stable adsorption site of the azobenzene molecule with $\rm CH_2S$ linker on Au(111) surface. (a):
initial position; (b): the optimized position.}
\end{figure}

In order to determine the preferred adsorption position for the azobenzene monolayer on the Au(111) surface,
three possible initial binding sites of both azobenzene isomers with the bottom $\rm CH_2S$ linker are
optimized on the Au(111) surface, the on-top site, the fcc- and hcp-bridge sites, as shown in Fig.\
\ref{fig:sites}. The calculated total energy and binding energy of an azobenzene molecule at different
adsorption sites are reported in Table \ref{tab:table}. For both \textit{trans} and \textit{cis}
configurations, we find the most stable adsorption position to be the fcc-bridge site. We also find the
\textit{trans} configuration to be more stable by about 0.6 eV than the \textit{cis} configuration, in good
agreement with the results from previous calculations \cite{McNellis:2009} and experiment
\cite{Schulze:1977}. Bader analysis based on the real-space charge density \cite{Henkelman:2006} for the
Au-AB shows a small amount of charge transfer from Au substrate to the molecule for all cases, as given in
Table \ref{tab:table}. The charge transferred to the molecule is primary localized at the $\rm CH_2S$ linker,
and not efficiently shared with the core part of the molecule. Hereafter we present only the results for
systems with the azobenzene monolayer chemisorbed on the fcc-bridge site of the Au(111) surface.

\begin{table}
\caption{\label{tab:table} Calculated total energy (in \rm eV per molecule), binding energy (in \rm eV per
molecule) and charge transfer (in $|e|$) of an azobenzene molecule on Au(111) surface at different adsorption
sites. The total-energy of Au/\textit{trans}AB at fcc-bridge site is set to be reference zero.}
\begin{ruledtabular}
\begin{tabular}{ccccccc}
isomer type &adsorption site & $E_{\rm total}$ & $E_b$  & charge transfer \\
\hline
trans & fcc-bridge & 0 & 2.084 & 0.07 \\
 & hcp-bridge & 0.110 & 1.974 & 0.07 \\
& on-top & 0.398 & 1.686 & 0.08 \\
cis & fcc-bridge & 0.635 & 2.112 & 0.08 \\
 & hcp-bridge & 0.700 & 2.037 & 0.07 \\
& on-top & 1.052 & 1.695 & 0.06 \\
\end{tabular}
\end{ruledtabular}
\end{table}

\begin{figure}
{\includegraphics[width=7.2cm]{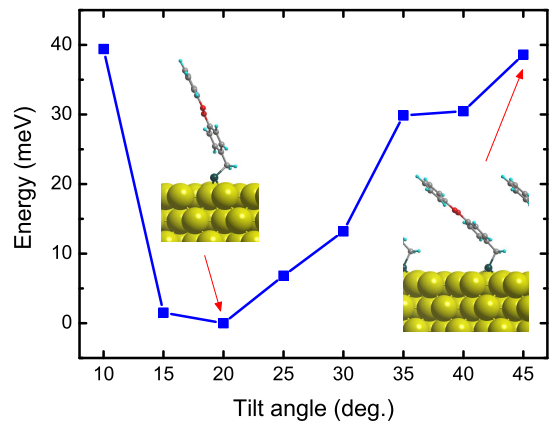}} \caption{\label{fig:angle}(Color online). Total energy as a
function of tilt angle for \textit{trans} azobenzene molecules chemisorbed on Au(111) surface. The left and
right insets are side views of the corresponding optimized structures with a tilt angle of
$20\textordmasculine$ and $45\textordmasculine$, respectively.}
\end{figure}

We also optimize the tilt angle of the azobenzene monolayer on Au(111) surface. In Fig.\ \ref{fig:angle} we
show the calculated total energy of a \textit{trans} azobenzene monolayer on Au(111) as a function of tilt
angle ranging from $10\textordmasculine$ to $45\textordmasculine$. At every single energy point in Fig.\
\ref{fig:angle}, the monolayer structure is optimized by performing geometry relaxation with an initially
given tilt angle. Within this angle range, we find that the lowest energy structure is the
$20\textordmasculine$ tilted molecular chemisorbed on the Au(111) surface. We expect the same finding for the
\textit{cis} configuration since the \textit{trans} to \textit{cis} transformation only change the geometry
of top half of the azobenzene molecule. Therefore, for the rest of the study we choose azobenzene monolayer
with tilt angle of $20\textordmasculine$ as the equilibrium structure for both \textit{trans} and
\textit{cis} configurations, as shown in Fig.\ \ref{fig:monolayer}.

However, it should be noted that in Fig.\ \ref{fig:angle} the calculated total energies of these systems with
different tilt angles are very close. In experiments, the tilt angle of molecules in the monolayer is usually
determined by other factors, such as ambient temperature and monolayer packing density. It has been found
that the molecular monolayer undergoes a phase transition from a tilted structure to a vertical structure at
room temperature \cite{Alkis:2007}. Also, in densely packed monolayers at full coverage, monolayers with a
small tilt angle can be generally expected, with molecules standing with their long molecular axes close to
the surface normal.

\subsection{\label{sec:33}Electronic structure}

\begin{figure}
{\includegraphics[width=8cm]{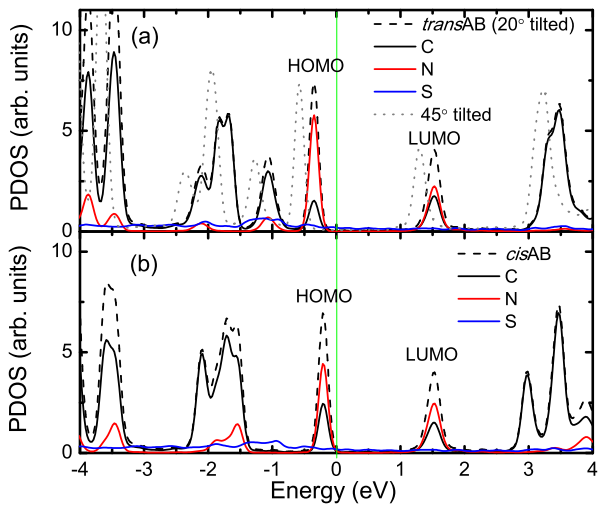}} \caption{\label{fig:pdos}(Color online). Projected density of
states of the molecule and various atoms of \textit{trans} (a) and \textit{cis} (b) azobenzene monolayer
chemisorbed on Au(111) surface. For \textit{trans} configuration, additionally shown is the PDOS of the
molecule with a tilt angle of $45\textordmasculine$ on the Au surface. Vertical green line represent the
Fermi energy which is at 0 eV.}
\end{figure}

Figure \ \ref{fig:pdos} analyzes the interaction between the azobenzene and Au(111) surface in terms of the
density of states projected onto the atoms (PDOS) in the molecule as well as the sulfur linker. For both
\textit{trans} and \textit{cis} configurations, we see the two frontier orbital levels of azobenzene, highest
occupied molecular orbital (HOMO) and lowest unoccupied molecular orbital (LUMO), lie at each side of the Au
Fermi level. Unlike the other molecular orbital levels nearby, the HOMO and LUMO contain large contributions
from the nitrogen pair of the azobenzene. The position of HOMO in the \textit{cis} configuration is closer to
the Au Fermi level than that in the \textit{trans} configuration. This is quite different compared to the
results from a study of a single azobenzene molecule on an Au(111) surface \cite{McNellis:2009}, in which the
downshift of molecular orbital level is more significant for the \textit{cis} isomer, resulting in the HOMO
of \textit{cis} azobenzene being away from the Au Fermi energy and its LUMO being very close to the Fermi
energy. This difference arises because the azobenzene molecules in the monolayer are standing upright (nearly
perpendicular to the surface), resulting in a weak interaction between the molecule and the bottom surface,
whereas a single azobenzene molecule orients parallel to the surface and thus the closer proximity between
the azo group (N pair) of the molecule and the Au surface creates a stronger interaction between them
\cite{McNellis:2009,Chapman:2010}. For \textit{trans} monolayer with a tilt angle of $45\textordmasculine$ on
the Au surface, we find essentially the same features in the PDOS as obtained for $20\textordmasculine$
tilted \textit{trans} monolayer, but all molecular orbital levels are roughly equally shifted by about 0.4 eV
to lower energies.

\section{\label{sec:4}Azobenzene monolayer junctions and transport properties}

\subsection{\label{sec:41}Structural models and electronic structures}

Once the equilibrium geometry of the Au-AB is obtained, we construct the corresponding molecular junction by
extending the bottom Au(111) substrate into a six atomic layer slab, and attaching a second Au(111) slab
consisting another six atomic Au layers on top of the azobenzene monolayer. To examine the effect of
monolayer-electrode contact, two types of interfacial contacts for the top electrode are studied: (1) the top
Au surface is covalently bonded to the azobenzen monolayer with $\rm CH_2S$ linkers same as those at the
bottom contact (hereafter we denote this type of junction as Au-AB-Au), and (2) the top contact is
physisorbed on the azobenzene monolayer surface without the $\rm CH_2S$ linkers (we denote this as
Au-AB$\mid$Au junction). We optimize the top interfacial contact by computing the total energies of the
system as a function of the distance between the left and the right electrodes. Every single energy point is
calculated by performing geometry optimization with a constrained electrode-electrode separation. Therefore
the equilibrium geometry is obtained as the distance at which the total energy is minimal. Due to the failure
of DFT GGA exchange correlation functionals to account for vdW interactions, we optimize the
electrode-electrode separation using vdW density functionals optB88-vdW \cite{Klimes:2011} for junctions with
physisorbed contacts. Fig.\ \ref{fig:junction} (a) and (b) shows the optimized monolayer junction geometries
with equilibrium electrode-electrode distance for \textit{trans} and \textit{cis} configurations,
respectively.

\begin{figure}
{\includegraphics[width=8cm]{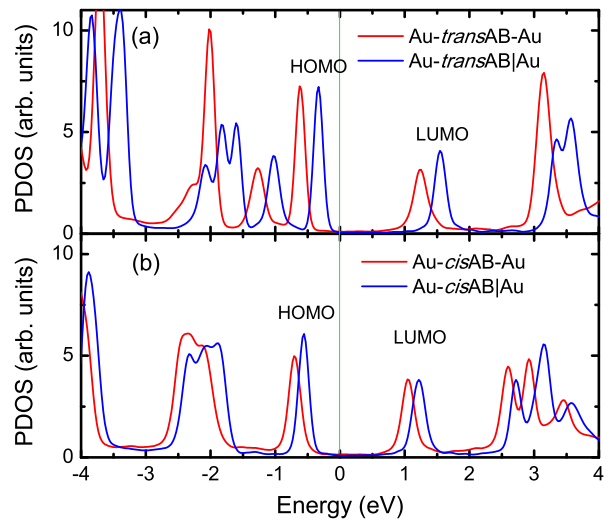}} \caption{\label{fig:pdos2}(Color online). Projected density of
states of the molecule in \textit{trans} (a) and \textit{cis} (b) azobenzene monolayer junctions.}
\end{figure}

We show the PDOS of the azobenzene molecules in the monolayer junctions in Fig.\ \ref{fig:pdos2}. As compared
to the PDOS of the bare monolayer on Au(111) surface, it can be clearly seen that all molecular orbital
levels, including the HOMO and the LUMO, are shifted to lower energies with the presence of a top
monolayer-electrode contact. In Fig.\ \ref{fig:pdos2} (a), the downshift of molecule orbital levels in the
Au-\textit{trans}AB$\mid$Au junction is only minimal, as compared to Fig.\ \ref{fig:pdos} (a), primarily
because of a weak bonding between the top electrode and the \textit{trans}AB molecule. In contrast, with the
sulfur linkers in the top contact of the Au-\textit{trans}AB-Au junction, a stronger monolayer-electrode
bonding significantly lower all molecular orbital levels in energies by about 0.3 eV. The downshift is even
more significant in the cases of \textit{cis} configurations as shown in Fig.\ \ref{fig:pdos2} (b). All
\textit{cis} azobenzene orbital levels shift to lower energies by about 0.4 eV in Au-\textit{cis}AB$\mid$Au
and 0.6 eV in Au-\textit{cis}AB-Au, compared to Fig.\ \ref{fig:pdos} (b).

\subsection{\label{sec:42}Transmission functions at zero-bias}

\begin{figure*}
{\includegraphics[width=16cm]{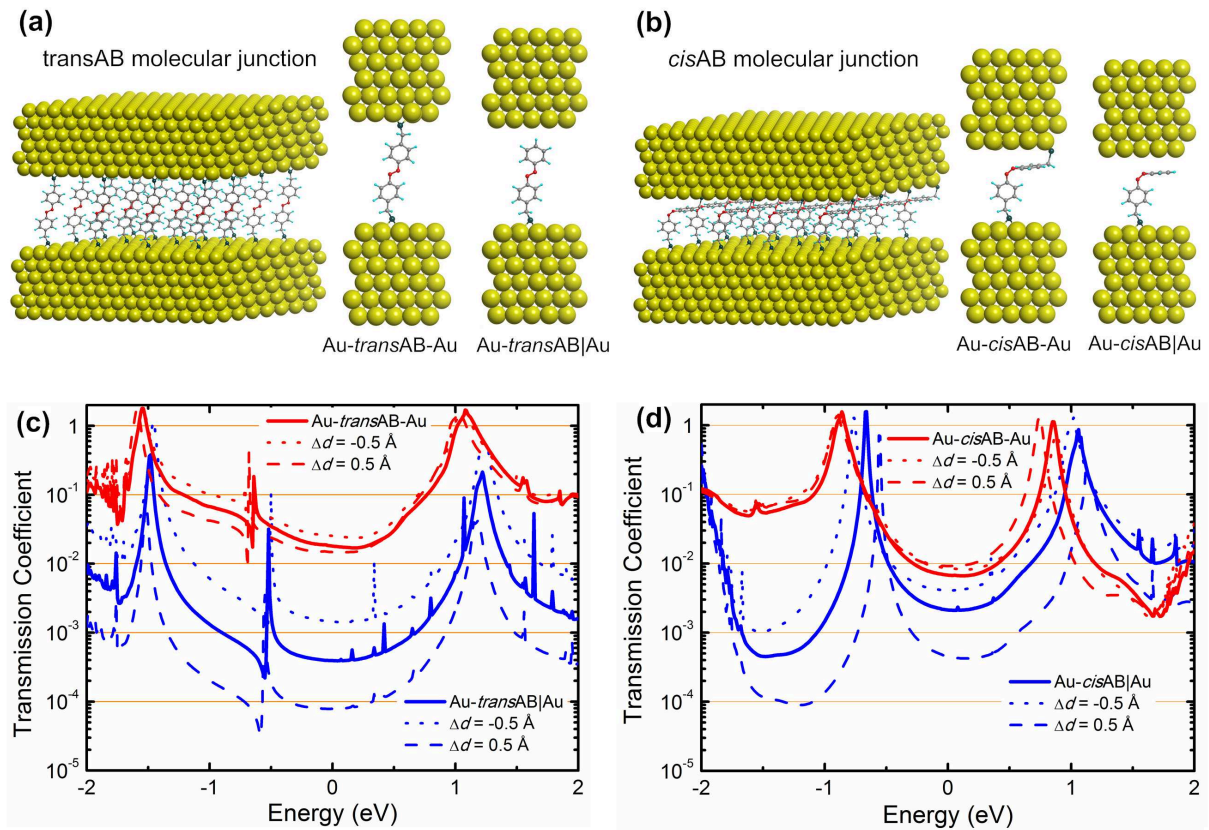}} \caption{\label{fig:junction}(Color online). Equilibrium
geometries (a-b) and zero-bias electron transmission functions (c-d) of \textit{trans} and \textit{cis}
monolayer junctions with two types of top contact. In panels (c) and (d) the y-axis is in logarithm scale,
and the Fermi level is shifted to 0. $\Delta d$ represent the change in the electrode-electrode separation
for a junction with its geometry slightly away from the equilibrium structure.}
\end{figure*}

In Fig.\ \ref{fig:junction} (c) and (d) we show the calculated transmission functions at zero-bias for
azobenzene monolayer junctions with \textit{trans} and \textit{cis} configurations, respectively. For the
Au-\textit{trans}AB-Au junction with chemisorbed top contact, transmission function has three broad resonance
peaks with near perfect transmission amplitudes shown in the given range of energy from -2 eV to 2 eV related
to the Fermi energy of the Au electrodes: two are below the Fermi energy (about -1.6 eV and -0.7 eV), and the
third is above it (about 1.1 eV). These three transmission peaks can be directly associated to the frontier
molecular orbitals HOMO-1, HOMO and LUMO, respectively, despite some minor changes in the energy as compared
to the PDOS in Fig.\ \ref{fig:pdos2} due to the periodic boundary condition used in the supercell calculation
for PDOS. The transmission peak associated with the HOMO of \textit{trans} azobenzene is relatively narrower
in width and lower in height because of the fact that the HOMO is more localized in the N pair of the
molecule and thus contributes less to the electron transmission. These resonance peaks in the transmission
function are away from the Au Fermi level, and zero-bias conductance of the monolayer junction is primarily
dominated by non-resonance electron tunneling at the Fermi level, with a transmission coefficient
$T\approx0.02$. For the Au-\textit{trans}AB$\mid$Au junction with a physisorbed top contact, we find that the
calculated transmission function is significantly decreased by about two orders of magnitudes as compared to
that of Au-\textit{trans}AB-Au junction, and, at the Fermi level, $T\approx0.0004$. The resonance
transmission peaks of Au-\textit{trans}AB$\mid$Au junction shift about 0.2 eV upward in the energy, which
follows closely the peaks in PDOS as shown in Fig.\ \ref{fig:pdos} (a). Besides three major peaks, there are
several small resonance peaks appearing in the transmission function of Au-\textit{trans}AB$\mid$Au around 0
eV to 1.2 eV which are induced by the vacuum gap in the top contact. Their relatively smaller amplitudes
provides only finite contribution to the tunneling current under certain bias voltages, which we will discuss
later in the next section.

For monolayer junctions with \textit{cis} configuration, however, removing the $\rm CH_2S$ linkers in the top
contact leads to a relatively much smaller decrease in transmission function near the Fermi energy, as shown
in Fig.\ \ref{fig:junction} (d). The transmission functions at the Fermi level for junctions
Au-\textit{cis}AB-Au and Au-\textit{cis}AB$\mid$Au are $T\approx0.006$ and 0.002, respectively. For both
Au-\textit{cis}AB-Au and Au-\textit{cis}AB$\mid$Au, within the given range of energy two resonance peaks can
be found in the transmission functions, which are associated with the \textit{cis} azobenzene HOMO and LUMO.
They are also away from the Au Fermi level, and the peaks of Au-\textit{cis}AB-Au junction are lower in
energy than that of Au-\textit{cis}AB$\mid$Au, corresponding to the PDOS peaks as shown in Fig.\
\ref{fig:pdos} (b).

The above calculated zero-bias transmission function cannot be simply studied by its relation to the
electrode-electrode separation or the azobenzene molecular length. Instead, we can describe the total
transmission function that reflects the efficiency of electronic transport from one electrode to the other
through the azobenzene monolayer junction as
\begin{equation}
T=T_{\rm bc}\cdot T^{\rm p1}_{\rm mol}\cdot T^{\rm p2}_{\rm mol}\cdot T_{\rm tc} \label{transmission}
\end{equation}
where $T_{\rm bc}$ and $T_{\rm tc}$ give the efficiency of electron transport across the bottom and top
contacts, and $T^{\rm p1}_{\rm mol}$ and $T^{\rm p2}_{\rm mol}$ reflect the electron transport through the
first (bottom half) and second (top half) parts of the azobenzene molecule, respectively. From Fig.\
\ref{fig:junction} (a) and (b), it is safe to assume that the first term $T_{\rm bc}$ in the above
transmission equation is the same for all four junctions. However, the other three terms, $T^{\rm p1}_{\rm
mol}$, $T^{\rm p2}_{\rm mol}$ and $T_{\rm tc}$, are quite different from case to case, resulting in different
total transmission coefficients at the Fermi energy for the above junctions. In the case of the symmetric
Au-\textit{trans}AB-Au junction, we have $T^{\rm p1}_{\rm mol}=T^{\rm p2}_{\rm mol}$ and $T_{\rm bc}=T_{\rm
tc}$, and the junction has a large transmission coefficient at the Fermi level, $T\approx0.02$. The absence
of the $\rm CH_2S$ linker in the Au-\textit{trans}AB$\mid$Au junction causes a large decrease in $T_{tc}$,
and reduces the total transmission coefficient to $T\approx0.0004$. One may expect such a large change in the
transmission term $T_{\rm tc}$ to also apply to the \textit{cis} configuration junction. However, as shown in
Fig.\ \ref{fig:junction} (d), the total transmission coefficient for Au-\textit{cis}AB$\mid$Au
($T\approx0.002$) is only about three times smaller than that for Au-\textit{cis}AB-Au ($T\approx0.006$).
This clearly indicate that, in the Au-\textit{cis}AB$\mid$Au junction, the transmitted electron travel
directly from the bottom contact to the top contact through only the first (bottom) part of the \textit{cis}
azobenzene molecule, and avoid the second (top) part of the molecule. Thus for the Au-\textit{cis}AB$\mid$Au
junction the effective tunneling pathway is greatly reduced, resulting a total transmission function
$T=T_{\rm bc}\cdot T^{\rm p1}_{\rm mol}\cdot T_{\rm tc}$. Note this is equal to Eq.\ \ref{transmission} with
term $T^{\rm p2}_{\rm mol}=1$ that compensates the smaller term $T_{\rm tc}$ of the Au-\textit{cis}AB$\mid$Au
junction, resulting in a smaller difference in the total transmission coefficients between the
Au-\textit{cis}AB$\mid$Au and Au-\textit{cis}AB-Au junctions as compared to that of the junctions in the
\textit{trans} configuration.

It should be also noted that the transmission terms $T^{\rm p1}_{\rm mol}$ and $T^{\rm p2}_{\rm mol}$ are
also different between junctions with \textit{trans} and \textit{cis} configurations. Non-resonant electron
transmission through a molecule $T_{\rm mol}$ can be empirically expressed as $T_{\rm mol}=exp(-\beta l)$,
where $l$ is the length of the molecule and represents the width of effective tunneling barrier, and $\beta$
is the tunneling decay factor given by $\beta=(2\sqrt{2m^*\phi}\alpha)/\hbar$ where $\phi$ is the barrier
height for tunneling that is determined by the frontier molecular orbital level related to the Fermi energy,
$\phi=E_{\rm F}-E_{\rm MO}$, $m^*$ is the effective mass and $\alpha$ is the shape constant of the barrier.
For junctions with chemisorbed top contact, at the Fermi energy the total transmission ratio between
\textit{trans} and \textit{cis} configurations is $t=T^{trans}_{\rm mol}/T^{cis}_{\rm mol}={\rm
exp}[(\beta_{cis}-\beta_{trans})2l]$. Giving $t\approx3.3$ from Fig.\ \ref{fig:junction} and $l\approx4.5
\AA$ estimated from the length of each half of the azobenzene molecule (remain unchanged upon switching), we
have $\beta_{cis}-\beta_{trans}=0.13 \AA^{-1}$. For junctions with physisorbed top contact, from the above
discussions we know that in the Au-\textit{trans}AB$\mid$Au the effective barrier width for the expression of
$T^{trans}_{\rm mol}$ is $2l$ and that in Au-\textit{cis}AB$\mid$Au is $l$, thus the total transmission ratio
between these two junctions at the Fermi energy should be $t'={\rm exp}[(\beta_{cis}l-\beta_{trans}2l)]$.
Giving $t'\approx0.2$ from Fig.\ \ref{fig:junction}, we get the tunneling decay factors $\beta_{trans}=0.49
\AA^{-1}$ and $\beta_{cis}=0.62 \AA^-1$. Both of them are in good agreement with the literature values
($\beta \approx 0.4-0.6 \AA^{-1}$) commonly obtained for short $\pi$-conjugated molecules
\cite{Salomon:2003}. A possible cause for a smaller tunneling decay factor $\beta_{trans}$ than $\beta_{cis}$
 could be the fact that the HOMO level of junction in the \textit{trans} configuration lie closer to the Fermi level
than the HOMO level of the junction in the \textit{cis} configuration, as shown in Fig.\ \ref{fig:pdos2} and
Fig.\ \ref{fig:junction}, resulting in a lower tunneling barrier, $\phi$. The barrier shape constant $\alpha$
could also be affected after the switching of azobenzene monolayer from \textit{trans} to \textit{cis} form.

In experiments, the monolayer-electrode contact of the junction may not be exactly at its equilibrium
position, but rather be slightly stretched or compressed. To simulate this situation, we calculate the
transport properties of junctions with an increased (or decreased) electrode-electrode separation $\Delta
d=0.5 \AA$ (or $\Delta d=-0.5 \AA$) with regard to its equilibrium value ($\Delta d=0$) for both
\textit{trans} and \textit{cis} monolayer junctions. The transmission results for stretched and compressed
junctions are also shown in Fig.\ \ref{fig:junction}. We find that the conductance of the Au-AB-Au junction
with chemisobed top contact is insensitive to the change of $\Delta d$ for both \textit{trans} and
\textit{cis} configurations, where as for Au-AB$\mid$Au junctions with physisobed top contact, a small change
in the electrode-electrode separation in either direction changes the conductance significantly. This
demonstrates that, with a physisorbed top contact, the total transmission function of the monolayer junction
is highly related to the $T_{\rm tc}$ (contribution from the contact) which is very sensitive to the
monolayer-electrode contact distance. In the case of Au-\textit{cis}AB$\mid$Au junctions, we also notice that
the two peaks in the transmission function at about -0.7 eV and 1.1 eV, which are associated with the HOMO
and the LUMO, move up in energy with increased electrode-electrode separation. This phenomenon has also been
noticed for other stretched molecular junctions \cite{Xue:2003}, in which the shift of HOMO resonance in the
transmission function toward the Fermi level increases the transmission coefficient at the Fermi energy.
However, here the decrease in the transmission coefficient at the Fermi level, due to the stretching of the
junction, is much more significant. The transmission peak associated with the \textit{cis} HOMO is still more
than 0.5 eV away from the Fermi level with $\Delta d=0.5{\AA}$, making the decrease in low bias conductance
unlikely to be compensated. In contrast, for stretched and compressed Au-\textit{trans}AB$\mid$Au junctions
we do not see such obvious change in the positions of transmission resonances.

\begin{figure}
{\includegraphics[width=7.5cm]{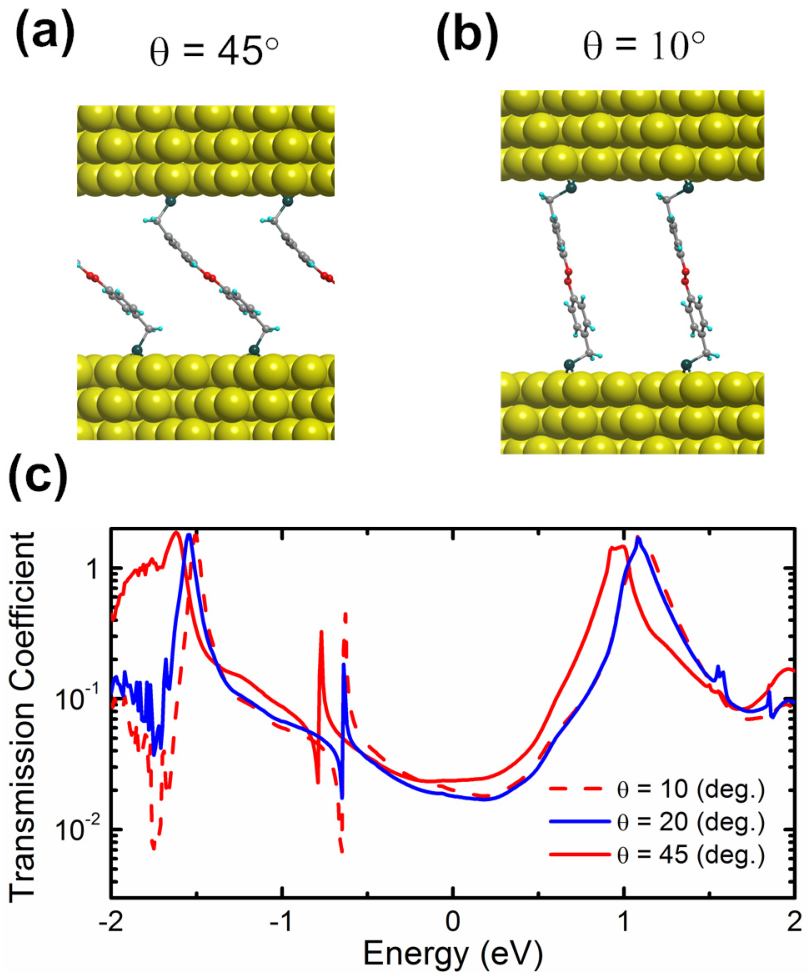}} \caption{\label{fig:angle2}(Color online). Equilibrium
geometries (a-b) and zero-bias transmission functions for Au-\textit{trans}AB-Au junctions with a tilt angle
$\theta$ of $45\textordmasculine$ and $10\textordmasculine$.}
\end{figure}

In order to understand the possible influences of the monolayer tilt angle $\theta$ on the transport
properties, we carry out further calculations for Au-\textit{trans}AB-Au junction with different $\theta$.
The optimized junction structures with $45\textordmasculine$ or $10\textordmasculine$ tilted \textit{trans}
monolayer are shown in Fig.\ \ref{fig:angle2} (a) and (b), respectively. A large change in the
electrode-electrode separation occurs between these two cases ($\Delta_d=2.5 \AA$). However, as shown in
Fig.\ \ref{fig:angle2} (c), for junctions with a $45\textordmasculine$ or $10\textordmasculine$ tilted
\textit{trans} monolayer, we find essentially the same features in the transmission functions as those
obtained for the optimized geometry with $\theta=20\textordmasculine$, yet with a small change in position of
transmission resonance peaks associated with the \textit{trans} azobenzene molecular orbital levels for
$\theta=45\textordmasculine$. The transmission at the Fermi level for $45\textordmasculine$ tilted junction
is only slightly increased as compared to the junctions with small tilt angles. The source of this increase
is the downward shifting of the broad peak in the transmission function associated with the \textit{trans}
LUMO. The shifting of molecular frontier levels resulting from the change in tilt angle may be resulted from
a change in the effective dipole moment as well as the work-function at the molecule-metal interfaces
\cite{Renzi:2005}. Similar phenomena have also been noticed in tilted alkanethiol monolayers
\cite{Frederiksen:2009}, in which the tilt brings the HOMO resonance closer to the Fermi energy.
Nevertheless, we conclude that for dilute monolayer junctions without intermolecular contribution, the
molecular tilt has very little effect on its conductance since the effective tunneling pathway through the
molecule remains unchanged upon tilting.

\subsection{\label{sec:43}Current-voltage characteristics}

\begin{figure}
{\includegraphics[width=7.5cm]{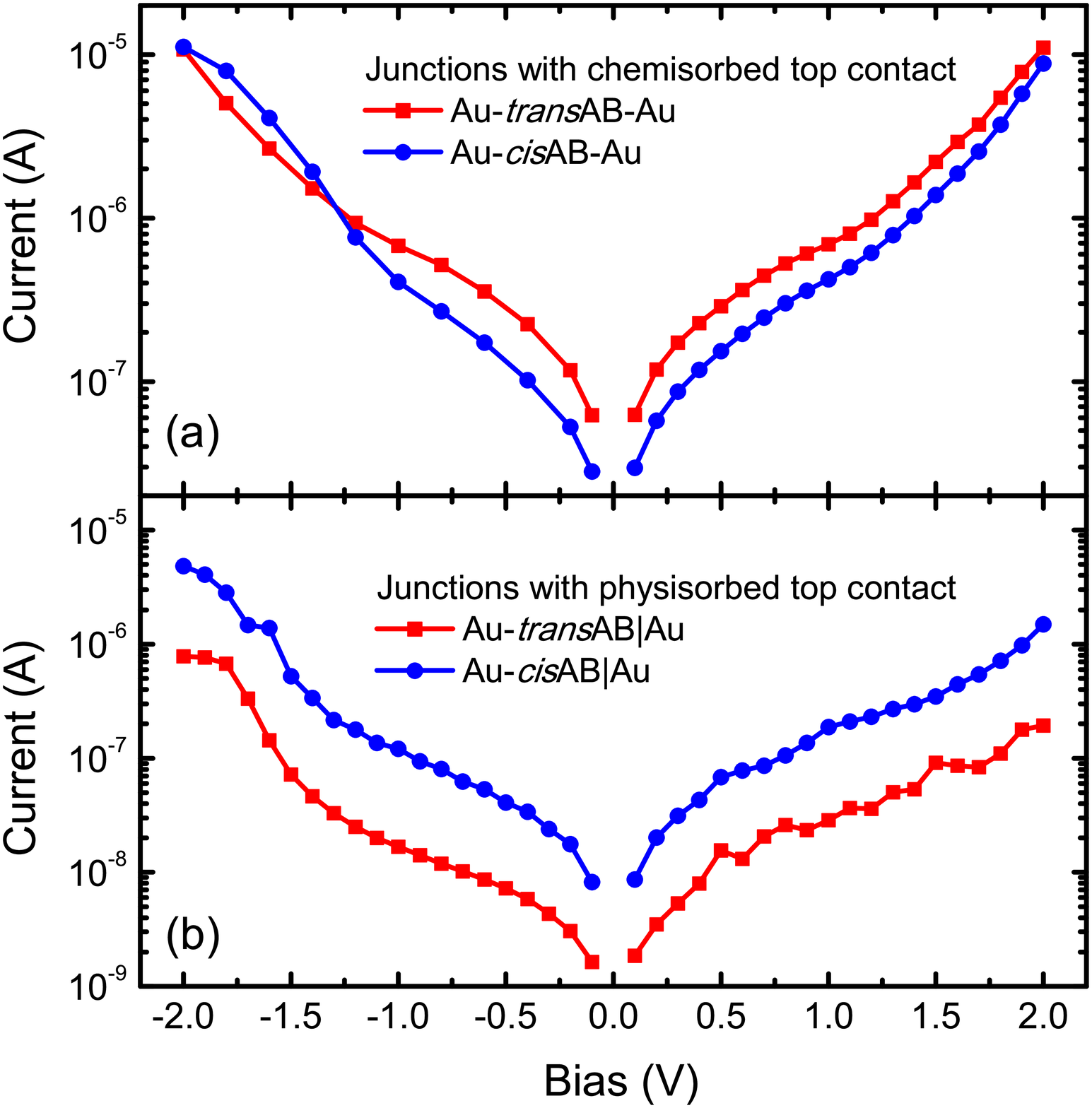}} \caption{\label{fig:current}(Color online). Current voltage
characteristics of Au-AB-Au (a) and Au-AB$\mid$Au junctions with \textit{trans} and \textit{cis}
configurations.}
\end{figure}

The self-consistently calculated current-voltage characteristics (I-V) of Au-AB-Au and Au-AB$\mid$Au
junctions are depicted in Fig.\ \ref{fig:current} (a) and (b), respectively. For the Au-\textit{trans}AB-Au
junction, a highly symmetric I-V curve arises as expected with two symmetric monolayer-electrode contacts,
and its current is larger than that of the Au-\textit{cis}AB-Au junction in small bias region, consistent
with the zero-bias transmission functions. Thus, with chemisorbed top contact, the two isomers realize
different conductance states of the monolayer junction corresponding to ``ON'' (\textit{trans} configuration)
and ``OFF'' (\textit{cis} configuration). The current of the Au-\textit{trans}AB-Au junction becomes smaller
than that of the Au-\textit{cis}AB-Au only in the region of large negative bias voltage (-1.2 V to -2.0 V).
This can be attributed to the fact that the Au-\textit{cis}AB-Au junction is asymmetric in geometry,
resulting in an asymmetric I-V curve which has a slightly larger current in the large negative bias region.
Such asymmetric behavior is also shown in the I-V curves for junctions Au-\textit{trans}AB$\mid$Au and
Au-\textit{cis}AB$\mid$Au, with asymmetric top (physisorbed) and bottom (chemisorbed) contacts. Nonetheless,
for junctions with a physisorbed top contact, the \textit{cis} configuration (``ON'' state) exhibits a larger
current for any given voltage than that of the \textit{trans} configuration (``OFF'' state), and the ON/OFF
ratio of the current is more pronounced than that of the junctions with chemisorbed top contacts. These
theoretical results for junctions with a physisorbed top contact are, therefore, in reasonable qualitative
consistency with the reported measurements \cite{Ferri:2008,Mativetsky:2008,Smaali:2010}. A quantitative
matching of the ON/OFF ratio to the experiments is not expected because the length of the molecules in
experiments are much longer than those in our study.

In Fig.\ \ref{fig:current} (b), we find the current of the Au-\textit{trans}AB$\mid$Au junction oscillates as
a function of positive bias voltage, which correspond to the small resonance peaks found in the transmission
function above the Fermi level as shown in Fig.\ \ref{fig:junction} (c). As we increase the voltage, these
resonance peaks sequentially join the bias window and contribute to the current, resulting in resonant
tunneling current at certain bias voltages. However, these resonances can only provide finite contribution to
the current in the \textit{trans} configuration and the current ratio between the \textit{trans} and
\textit{cis} configurations is almost unaffected.

\subsection{\label{sec:44}Effect of molecular packing density}

\begin{figure}
{\includegraphics[width=7.4cm]{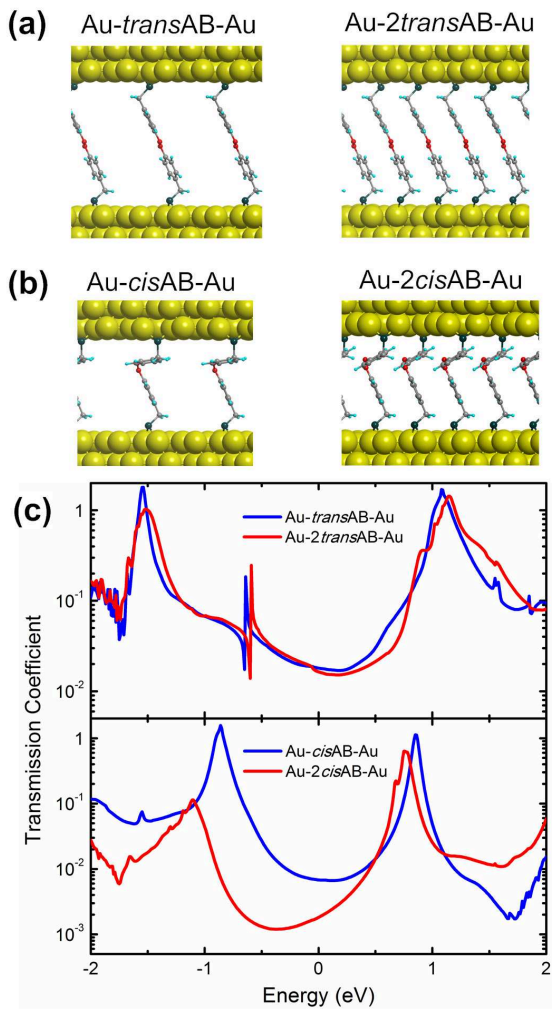}} \caption{\label{fig:density}(Color online). Equilibrium
geometries (a-b) and zero-bias transmission functions (c) of azobenzene monolayer junctions with different
packing density. The calculated transmission coefficients for junctions with high monolayer packing density
are divided by two (number of azobenzene molecules in one cell) in order to compare with the results with low
packing density.}
\end{figure}

The above calculations are based on a diluted monolayer structure with a low packing density, in which the
intermolecular bonding between molecules in the monolayer can be considered negligible and junction
transport properties are primarily dominated by the intramolecular contribution. To inquire into effect of
intermolecular interaction on the transport, Au-AB-Au junctions with a higher monolayer packing density are
also considered, in which the molecules are positioned closer to each other to increase the molecular
interactions. In this case for each junction we have two azobenzene molecules per
$(\sqrt{3}\times\sqrt{3})R30\textordmasculine$ Au(111) unit cell, and the packing area corresponds to 45.4
$\AA$ per molecule. The optimized junction structures for \textit{trans} and \textit{cis} configurations are
shown in Fig.\ \ref{fig:density} (a) and (b), and we denote them as Au-2\textit{trans}AB-Au and
Au-2\textit{cis}AB-Au, respectively. For the \textit{cis} configuration, as compared to the
Au-\textit{cis}AB-Au junction with a low packing density, we see that the azobenzene molecules in the
monolayer of the Au-2\textit{cis}AB-Au junction become distorted, so that the top half of the molecule is no
longer parallel to the top Au surface but instead rotated relative to the Au surface, resulting from an
increased intermolecular interaction with decreased spacing between the molecules.

The calculated zero-bias transmission functions are shown in Fig.\ \ref{fig:density} (c). For the
\textit{trans} configuration, doubling the packing density produces only negligible changes in transmission
function. However, for the \textit{cis} configuration, the junction's transmission coefficients are
significantly reduced especially in the energy region around the Fermi level. The decrease in conductance by
increasing the packing density has also been observed in other monolayer junctions \cite{Agapito:2008}, but
the reason has been attributed to a shift of the LUMO resonance peak in the transmission function to a higher
energy away from the Fermi level. Our result differ from the Ref.\ \onlinecite{Agapito:2008} in that the LUMO
peak in the transmission function of the Au-2\textit{cis}AB-Au junction slightly shifts to a lower energy
which is closer to the Fermi level as compared to the Au-\textit{cis}AB-Au. Nevertheless, the transmission
coefficients for the HOMO resonance peak and for the energy region around the Fermi level are decreased by
about one order of magnitude. The considerable decrease in the transmission function can be attributed to the
geometry distortion in the \textit{cis} azobenzene monolayer which is caused by intermolecular interactions
with the dense packing, as shown in Fig.\ \ref{fig:density} (b). Our results show clearly that interaction
among the azobenzene molecules of the monolayer junction can induce large changes in transport in the
\textit{cis} configuration, hence suggesting that higher conductance ON/OFF ratio can be achieved in densely
packed azobenzene monolayer junctions with chemisorbed monolayer-electrode contact. For junctions with a
physisorbed contact, since the ``ON'' state is associated with the \textit{cis} configuration, we expect that
increasing the monolayer packing density could decrease the conductance ON/OFF ratio.

\subsection{\label{sec:45}Effect of top-contact geometry}

\begin{figure}
{\includegraphics[width=7.5cm]{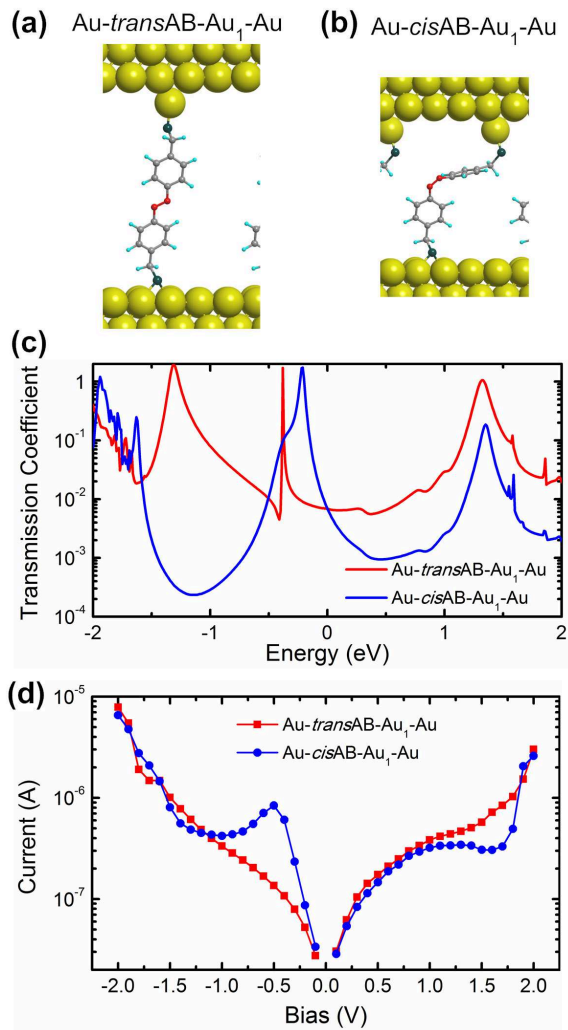}} \caption{\label{fig:Au1}(Color online). Equilibrium geometries
(a-b), zero-bias electron transmission coefficients (c) and current-voltage characteristics of Au-AB-$\rm
Au_1$-Au junctions.}
\end{figure}

Having addressed the transport properties of azobenzene molecular junctions with a perfect contact surface at
the monolayer-electrode interface, it is necessary to also study the consequences of non-ideal situations. To
investigate the role of contact atomic structure, we consider a simple but possible situation: the sulfur
atom bound to a single Au atom protruding from the surface of the top Au electrode, which is shown in Fig.\
\ref{fig:Au1} (a) and (b) for \textit{trans} and \textit{cis} configurations, and we denote them as denoted
as Au-AB-$\rm Au_1$-Au and Au-2\textit{cis}AB-Au, respectively. The purpose of this consideration is to
simulate possible situations in experiments in which the surface of top electrode may not be atomically flat
when contacting the molecules in the monolayer.

The calculated transmission functions are shown in Fig.\ \ref{fig:Au1} (c). We find that the additional Au
atom decreases the transmission coefficients at non-resonance tunneling regions in the transmission function
for both \textit{trans} and \textit{cis} configurations, as compared to Fig.\ \ref{fig:junction}. Moreover,
the resonance peaks in the transmission functions associated with the frontier molecular orbitals, all shift
up in energy at a noticeable change. The HOMO transmission resonance of Au-\textit{trans}AB-$\rm Au_1$-Au
junction shifts from -0.7 eV to -0.4 eV; while that of Au-\textit{cis}AB-Au junction shifts from -0.85 eV to
-0.2 eV, which is very close to the Fermi energy.

In Fig.\ \ref{fig:Au1} (d) we show the calculated current-voltage characteristics for Au-AB-$\rm Au_1$-Au
junctions. The most noticeable feature appearing in Fig.\ \ref{fig:Au1} (d) is the asymmetry in the cis
configuration. For a negative bias voltage below 1 eV, the current through the \textit{cis} configuration is
significantly higher than the current through the \textit{trans} configuration. As the negative bias is
further increased, the current through the \textit{cis} configuration decreases dramatically with increasing
voltage, resulting in a region of negative differential resistance (NDR). This NDR feature also appears in
the positive bias region (from 1.2 V to 1.7 V) in the \textit{cis} configuration, but is not present for any
bias region in the \textit{trans} configuration. In contrast to Au-\textit{cis}AB-$\rm Au_1$-Au, the $I$-$V$
characteristics of Au-\textit{trans}AB-$\rm Au_1$-Au junction are basically unchanged by the presence of the
additional Au atom as compared to Fig.\ \ref{fig:current} (a).

\begin{figure}
{\includegraphics[width=7.5cm]{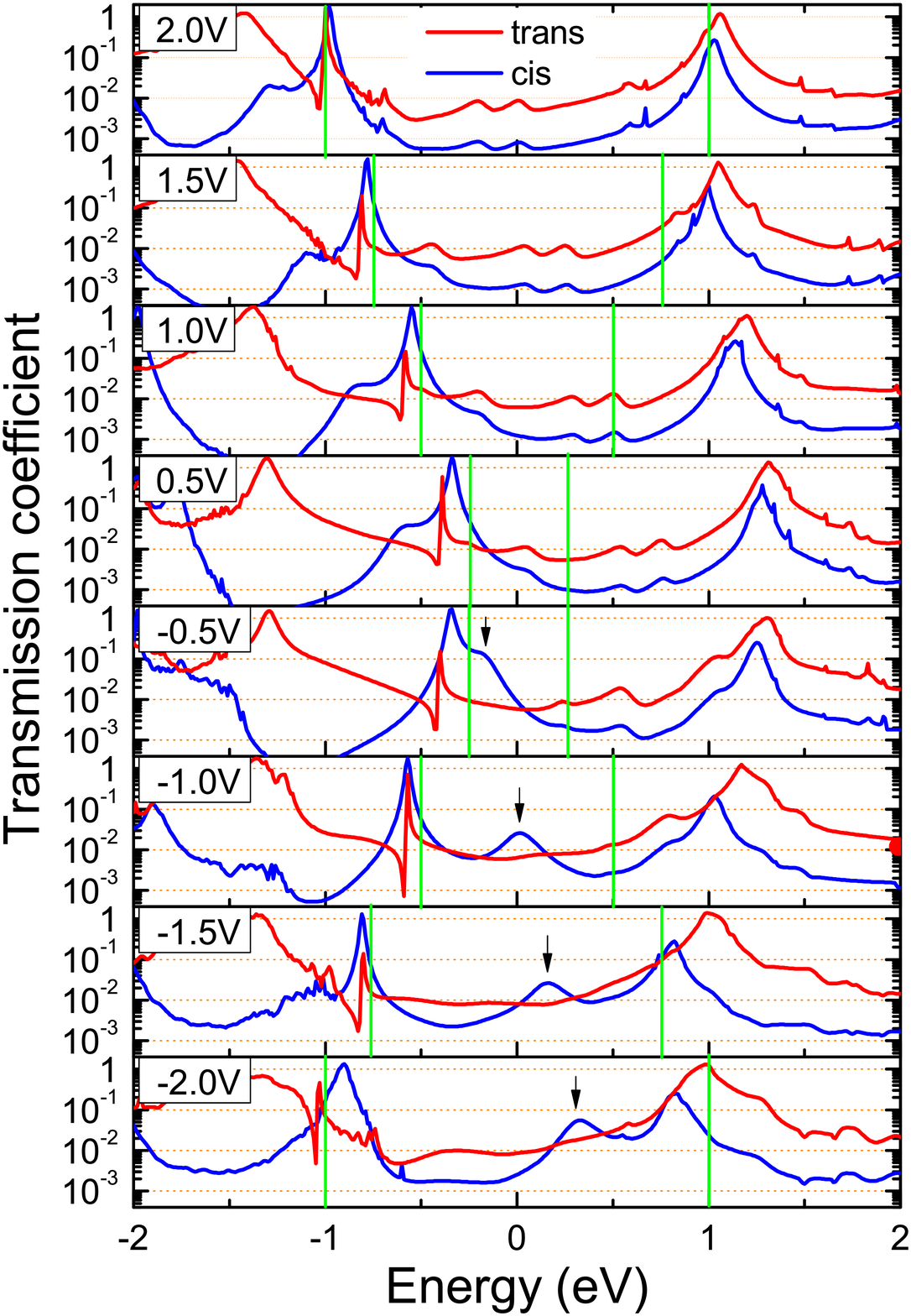}} \caption{\label{fig:T_bias}(Color online). Electron
transmission coefficients of Au-AB-$\rm Au_1$-Au junctions with \textit{trans} and \textit{cis}
configurations under various bias voltages from -2.0 to 2.0 V. The region between two green vertical lines
indicates the bias window. The arrow indicate the bias-induced transmission peak in the transmission function
in the \textit{cis} configuration.}
\end{figure}

To shed light on the origin of the NDR effect, we give the transmission function $T(E,V)$ at a series of
biases from $-2.0$ to 2.0 V at 0.5 V intervals as shown in Fig.\ \ref{fig:T_bias}. In the case of the
\textit{cis} configuration, a voltage of $-0.5$ V creates a bias-induced transmission peak at about $-0.2$ eV
(marked with an arrow in Fig.\ \ref{fig:T_bias}) adjacent to the resonance peak associated with molecular
LUMO. As the negative bias is further increased, this transmission peak, with a broad width, moves up in
energy and towards the molecular LUMO resonance. It dominates the transmission function within the bias
window, thus raising a much larger current than that at positive bias and resulting in a noticeable
rectification effect for Au-\textit{cis}AB-$\rm Au_1$-Au junction in Fig.\ \ref{fig:Au1} (c). In the case of
\textit{trans} configuration, however, only several bias-induced transmission peaks exist at positive bias,
and all of them have much lower amplitudes and shorter width leading to only small current contributions.
Correspondingly, in Fig.\ \ref{fig:Au1} (c) the rectification behavior in the I-V curve of
Au-\textit{trans}AB-$\rm Au_1$-Au junction is not obvious and there is no NDR effect. This demonstrate that
the azobenezene monolayer junction can be used for photo-controlled molecular rectifier with proper
engineering of its contact geometry.

It should also be noted that for both \textit{trans} and \textit{cis} configurations, the bias drives the
LUMO resonance peak away from the Fermi level, so its contribution to the current becomes important only at a
sufficiently large bias (e.g. $V\approx 2$ V) which should be comparable to the HOMO-LUMO gap of the
molecule. This is also true for the calculated $I$-$V$ characteristics in Sec.\ \ref{sec:43}. Therefore, at
low bias the electronic transport of all the monolayer junctions in consideration is mainly facilitated by
non-resonant electron tunneling.


\section{\label{sec:5}Conclusion}

Our first-principles calculations indicate that monolayer-electrode bonding, intermolecular interaction, and
contact geometry all play very important roles in determining the conductance ratio of azobenene monolayer
junction between \textit{trans} and \textit{cis} configurations. With a strongly bonded top contact from
chemisorption, the zero-bias transmission of \textit{trans} monolayer is higher than that of the \textit{cis}
monolayer. Replacing it by a weakly bonded contact with physisorption will decrease the conductance by about
two orders of magnitudes for the \textit{trans} monolayer. However, in the case of the \textit{cis}
configuration, the physisorbed contact shortens the effective tunneling pathway, leading to a conductance
that is less sensitive to weak bonding. The mechanism are interpreted by applying a simple transmission model
to the calculated zero-bias transmission functions, which accounts for electron transmission through each
subunit of the molecular junction and provides a clear physical picture for understanding the junction. The
calculated current-voltage characteristics indicate that, for junctions with physisorbed top contact, the
"ON" state with larger current is associated with the \textit{cis} configuration, which are in agreement with
recent experiments. Our calculations demonstrate the intermolecular interaction in the \textit{trans}
configuration is negligible even in a densely packed monolayer; while for \textit{cis} configuration
increasing the monolayer density causes a considerable distortion in the monolayer structure, resulting a
decrease in its transmission function thus increasing the conductance ratio. We also find that the calculated
current-voltage characteristics for the \textit{trans} monolayer remain symmetric upon changing of the top
contact geometry. In contrast, a slightly modified contact geometry will affect the electric current through
the \textit{cis} monolayer, leading to a highly unsymmetrical current-voltage curve as well as a large
negative differential resistance behavior. These results suggest that the molecule-lead contact, the state of
the molecule, the morphology of the metal surface, and the packing density of the monolayer are all
parameters in the play. Our investigations thus will deepen understanding electron transport through
azobenzene monolayer junctions.

This work is supported by US/DOE/BES/DE-FG02-02ER45995. The authors acknowledge DOE/NERSC and UF-HPC centers
for providing computational resources.

\end{document}